\documentclass[aps,pre,showpacs,amsfonts,amssymb,amsmath,floatfix,twocolumn,
floats,nobalancelastpage]{revtex4}

\usepackage{graphicx}

\begin{document}          

\title{Nonuniversal effects in mixing correlated-growth processes 
with randomness: Interplay between bulk morphology and surface roughening}

\author{A. Kolakowska}
\email{alice@kolakowska.us}
\affiliation{Department of Physics, The University of Memphis, 
Memphis, TN 38152}
\author{M. A. Novotny}
\affiliation{Department of Physics and Astronomy, and Center for 
Computational Sciences, 
P.O. Box 5167, Mississippi State, MS 39762-5167}

\date{\today}

\begin{abstract}
To construct continuum stochastic growth equations for competitive 
nonequilibrium surface-growth processes of the type RD+X that mixes random 
deposition (RD) with a correlated-growth process X, we use a 
simplex decomposition of the height field.   
A distinction between growth processes X that do and do not 
create voids in the bulk leads to the definition of the 
{\it effective probability} $p_{\mathrm{eff}}$ of the process 
X that is a measurable property of the bulk morphology and 
depends on the {\it activation probability} $p$ of X in the 
competitive process RD+X. The bulk morphology is reflected 
in the surface roughening via {\it nonuniversal} prefactors 
in the universal scaling of the surface width that scales in 
$p_{\mathrm{eff}}$. The equation and the resulting scaling are 
derived for X in either a Kardar-Parisi-Zhang or Edwards-Wilkinson 
universality class in $(1+1)$ dimensions, and illustrated by 
an example of X being a ballistic deposition.  We obtain full 
data collapse on its corresponding universal scaling function 
for all $p \in (0;1]$. We outline the generalizations to $(1+n)$ 
dimensions and to many-component competitive growth processes.
\end{abstract}


\maketitle

\section{Introduction \label{intro}}

Many dynamical complex physical systems in nature are studied 
by their mapping onto a suitable nonequilibrium surface-growth problem.  
The dynamics of correlation buildup in these physical systems, 
and their other properties, can then be explored with the use of 
surface-growth methodologies. Numerous examples of such studies, 
experimental as well as theoretical and computational, come from a 
variety of fields such as tumor-growth processes \cite{BAS03} in cancer research,  
growth of cell colonies \cite{HMP+14} in biophysics, roughening of 
lipid bilayers \cite{VMM14} in soft-matter biomaterials,  
dynamics of combustion fronts \cite{BM05}, 
imbibition processes \cite{SMP+05}, 
film-growth processes \cite{DSR03},  
time-series and market price analyses 
in econo-physics \cite{Bal07}, and scalability and synchronization of 
parallel-computing system \cite{KNG+03,KN05} in computer science, to give 
representative examples.

Large-scale properties are described within a continuum model by 
universal stochastic growth equations and tested with simulation models. 
On the theory side, the trouble is that simple discrete models such as SOS 
(solid-on-solid)  
are often not adequate to reproduce the complex physics of observed surface 
phenomena, because they assume only one universal process alone being responsible for 
surface formation. Such an idealization does not reflect actual experimental 
settings where the observed surface phenomena may involve contributions from 
several universal processes. The continuum description of such a 
multicomponent growth process has not yet been developed. Such 
mixed-growth systems display many nontrivial properties 
\cite{PJ90,Aus93,WC93,Kor98,DK00,EC00,DM01,HMA01,HA01,CR02,Rei02,HA03,
KNV04,MBB04,Reis04,CPT04,BL05,IMA05,Reis06,OD+06,KNV06,HA06,KN10,
CP09,Kola14,BSR14}.

A representative example comes from an applied model in computer 
science \cite{KN05} when the asynchronous dynamics of 
conservative updates in a system of parallel processors 
is modeled as a virtual-time surface that represents nonequilibrium processes 
in this system. When the load per processor is minimal, this dynamics belongs to 
the Kardar-Parisi-Zhang (KPZ) universality class \cite{KTN+00}. 
However, when the load 
is increased to reflect real operations, the realistic dynamics is a competitive 
growth process that combines a universal KPZ process with a random 
deposition (RD) process, i.e., is of the type RD+KPZ \cite{KNV04}. 
Consequently, in order to fully understand the statistics of 
the updates and make quantitative predictions, 
it is important to know how the \textit{nonuniversal} properties of 
the multi-component processes affect the universal scaling of a RD+KPZ process. 
This is still an unsolved problem of nonequilibrium surface growth science. 
In applied modeling, even if not explicitly assumed, competitive dynamics 
naturally arises. Studying these systems should 
also contribute to the understanding   
of differences between the expected and the actual scaling of rough interfaces, 
often encountered both in simulations and in experiments.

By a competitive-growth process Y+X --- alternatively called a two-component 
system or a mixed-growth process --- we understand a dynamical process where 
process Y alternates with process X in accordance with the rule of the exclusive 
alternative: ``\textit{either} process Y (active with probability $q$) \textit{or} 
process X (active with probability $p$),'' is active.  
Here $q+p=1$, and Y belongs to a 
different universality class than X. It is understood that in competitive 
surface-growth processes an event on the surface is triggered by only one 
process at a time, even if both mechanisms X and Y are simultaneously 
present.

In this article we introduce a method by which a continuum stochastic 
growth equation can be constructed for competitive growth processes. 
We investigate a connection between surface roughening 
and the bulk morphology formed during the deposition in the competitive growth  
process RD+X, where X is a correlated growth process of 
universal dynamics different  
from RD. This connection has been already established in simulations of 
competitive growth models \cite{PJ90} and of binary growth of thin 
films \cite{DK00}, and for diffusion-limited-aggregation models \cite{Krug95}. 
A new aspect of our study is to provide a direct 
theoretical link between nonuniversal properties of process X, as read 
from the bulk, and the continuum equation that underlies the observed 
universal scaling laws for the competitive RD+X processes. 
In this work, we derive from first principles a continuum equation 
to show that its model-dependent coefficients do reflect the bulk structure. 
This will lead to a distinction between void-producing and simple desorption 
and adsorption processes. As discussed later, 
this division into subclasses is a {\it necessary} first step towards a  
theory of many-component processes. In particular, 
it explains variations in scale dilatations observed in RD+X 
models \cite{KNV04,KNV06,Reis06,HA06,HA03,HMA01,HA01,KN10,CP09,Kola14,BSR14}. 
In our analysis we use as an example the universal RD+KPZ growth process 
in $(1+1)$ dimensions and generalize our approach to other processes 
in $(1+n)$ dimensions.

This article is organized as follows:  
Scaling of the interface width in competitive RD+X models is outlined in 
Sec.~\ref{scale}, where we show that the full data collapse scaling can be obtained 
in a geometric scaling given by Eqs. (\ref{eq00a}) and (\ref{eq00b}). 
This type of scaling was heuristically proposed in Ref. \cite{CP09} and its 
explicit form was derived in Ref. \cite{Kola14}. Geometric scaling 
confirms that the RD+X systems are in the universality class of 
process X \cite{KNV06}, but such data collapse is not a dynamic 
finite-size scaling. In the remaining part of Sec.~\ref{scale} we focus 
on dynamic scaling that provides a connection with stochastic dynamics 
as described by a continuum-growth equation.
In Sec.~\ref{conteq}, where we define the adsorption-bulk-compact and the 
dense-or-lace-bulk processes, we use a concept of simplectic decomposition 
to derive from first principles the stochastic growth equation for simple 
RD+X processes.  Hence, we find a connection between the bulk morphology 
and the surface roughening for these processes. Results of Sec.~\ref{conteq} 
are discussed in Sec.~\ref{discuss}, where we demonstrate by examples 
that in RD+X processes the nonuniversal prefactors in Family-Vicsek 
universal scaling function are nontrivial and have connection with 
the bulk morphology. In Sec.~\ref{discuss} we also give the extension 
of the approach introduced in Sec.~\ref{conteq} to $(1+n)$ 
dimensional models of two-component processes, outline a possible 
generalization to many-component competitive processes, and discuss 
further developments.  
Conclusions are summarized in Sec.~\ref{conc}.

\section{Scaling of the interface width \label{scale}}

Time-evolution of the correlation length is reflected in the interface 
width $w(t)$ of the growing surface. Both the correlation length 
and $w(t)$ have the same scaling properties. 
In SOS models of surfaces growing on a substrate of $L$ sites, 
$w(t)$ is measured as: $\langle w^2(t) \rangle = 
\langle L^{-1} \sum_{k=1}^{k=L} [h_k(t)- \bar{h}(t)]^2 \rangle$, 
where $h_k(t)$ is the column height at site $k$ at time $t$, and 
$\bar{h}(t)$ is its mean over $L$ sites. The time $t$ is measured as a 
number of deposited monolayers. The angular brackets denote configurational 
averages. For brevity of notation, we set  
$w \equiv \sqrt{\langle w^2 \rangle}$.

\begin{figure}[t]
\includegraphics[width=8.0cm]{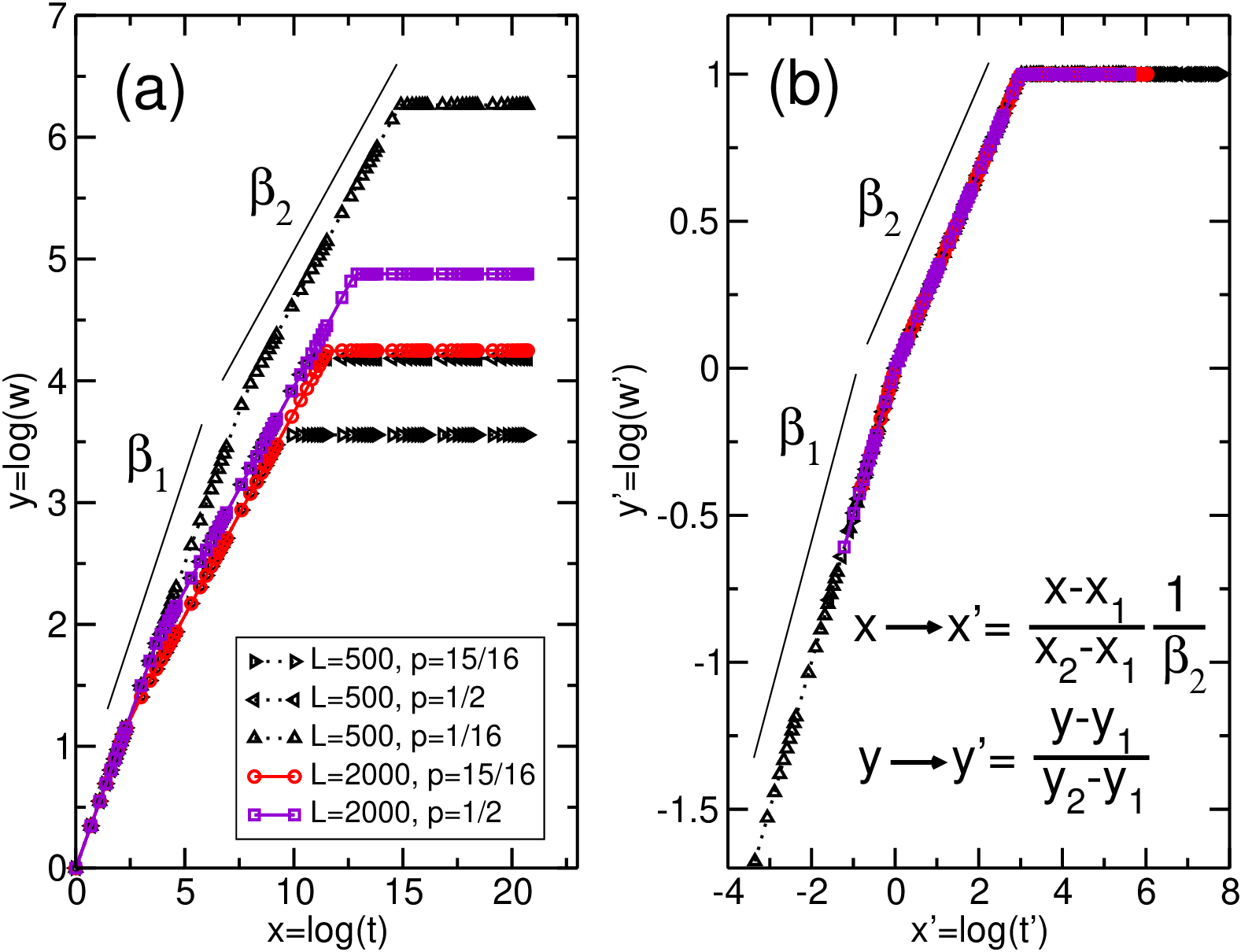}
\caption{\label{geometric_scaling}
(Color on line)
Typical time evolution of the interface width $w(t)$ in a competitive 
process RD+X for selected system sizes $L$ and activation probabilities 
$p$ of process X. 
(a) Synthetic data from Eq. (\ref{eq00}) when process X is in the KPZ 
universality class. (b) The collapse of the data in (a) that is 
produced by the geometric scaling $(x,y) \to (x',y')$ given by 
Eqs. (\ref{eq00a}) and (\ref{eq00b}) (shown in the figure). 
Here, $\beta_1=1/2$ and $\beta_2=1/3$ and $\log (\cdot) \equiv \log_{10} (\cdot)$.
}
\end{figure}

In competitive growth processes that mix correlated-growth process X 
with randomness, i.e., of the type RD+X, results of simulations 
with a flat-substrate initial condition at $t=0$  
can be summarized by the following two-parameter family of curves \cite{KNV04}:
\begin{equation}
\label{eq00}
w(t;L,p)=\left\{ \begin{array}
{r@ {\quad , \quad} l}
c_1 \sqrt{t} & t\in [0;t_1(p)] \\
c_2 t^{\beta_2} & t\in (t_1(p);t_2(L,p)) \\
c_3 L^{\alpha_2} & t\in [t_2(L,p);+\infty]  ,
\end{array} \right.
\end{equation}
where $p$ and $L$ are parameters, and 
$c_1$, $c_2$, and $c_3$ are constants.
The effect of the parameter $p$ on the time evolution of $w(t;L,p$$=$$1)$ 
is a nonuniversal dilatation of time and length scales, as discussed in 
Ref. \cite{KNV06}. Because of these dilatations the times $t_1(p)$, when 
the initial RD transients terminate, and the times $t_2(p)$, when the saturation 
phases begin, have different values for different curves $w(t;L,p)$. 
In Eq. (\ref{eq00}) $\beta_2$ and $\alpha_2$ are universal
scaling exponents (the growth and the roughness
exponents, respectively) characteristic of the universality
class of process $X$. For processes RD+KPZ 
(when $\alpha = \alpha_2 = 1/2$ and $\beta = \beta_2 = 1/3$) 
the family given by Eq. (\ref{eq00}) is illustrated in 
Fig. \ref{geometric_scaling}a, where $y=\log{w}$ is plotted 
versus $x=\log{t}$. In the $(x,y)$-plane, the family in Eq. (\ref{eq00}) 
can be collapsed onto one curve shown in Fig. \ref{geometric_scaling}b 
by means of the shift-and-scale operator 
$\hat{G}: (x,y) \longrightarrow (x',y')$ \cite{Kola14},  
\begin{eqnarray}
x & \to & x' = \frac{x-x_1}{x_2-x_1} \frac{1}{\beta_2} , \label{eq00a} \\
y & \to & y' = \frac{y-y_1}{y_2-y_1} , \label{eq00b}
\end{eqnarray}
where $x'=\log{t'}$ and $y'=\log{w'}$. 
In Eqs. (\ref{eq00a}) and (\ref{eq00b}) the pairs of numbers  
$x_1=\log{t_1(p)}$ and $y_1=\log{w_1(p)}$, and, 
$x_2=\log{t_2(L,p)}$ and $y_2=\log{w_2(L,p)}$, 
have different values for each curve $w(t;L,p)$. 
Explicitly, $w_1=w(t_1(p))$ and $w_2=w(t_2(L,p))$ 
by Eq. (\ref{eq00}). 
Operator $\hat{G}$ shifts all curves in Eq. (\ref{eq00}) to one position where 
all crossover points $(t_{\times},w_{\mathrm{sat}})$ to saturation are 
mapped onto one point $(1/\beta_2,1)$. Subsequently, $\hat{G}$  
scales the length $(x_2-x_1)\sqrt{1+1/\beta_2^2}$ of the correlated-growth 
phase in $(x,y)$-plane for each curve to the length of 
$\sqrt{1+1/\beta_2^2}$. The full data collapse obtained by $\hat{G}$ 
is possible because each curve in Eq. (\ref{eq00}) has one universal 
footprint, where the initial RD transient is followed by a 
specific universal correlation phase. Such a collapse of the data in the 
$(x,y)$-plane by geometric scaling is an illustration of the previously proven fact 
\cite{KNV06} that competitive-growth processes RD+X are in the 
universality class of process X. 
It must be stressed, however, that this geometric scaling expressed by $\hat{G}$ 
does not give the universal dynamic 
scaling function that would explain the universal shape of the curve 
in Fig. \ref{geometric_scaling}b in terms of finite-size scaling of the corresponding 
stochastic dynamics as described by the continuum model and, 
possibly, nonuniversal corrections to scaling when $p \ne 1$. 
Manipulation of Eqs. (\ref{eq00a}) and (\ref{eq00b}) to obtain 
explicitly $w(t)$ leads back to Eq. (\ref{eq00}); thus, 
Eqs. (\ref{eq00a}) and (\ref{eq00b}) do not contain any new physical 
information in addition to that already present in Eq. (\ref{eq00}). 
In summary to this point, the geometric scaling lacks  
physical meaning and does not connect with the Family-Vicsek dynamic scaling.

A dynamic scaling hypothesis for competitive RD+X processes \cite{KNV06} 
states that if a correlated growth X occurs with a constant probability 
$p$, its continuum equation must be invariant under the scaling
\begin{equation}
\label{scaling}
x \to x \, , \; h \to h/g(p) \, , \; t \to t/f(p) \, ,
\end{equation}
where $g(p)$ and $f(p)$ are arbitrary suitable functions of $p \in (0;1]$. 
This invariance implies that $f(p)=g^2(p)$. When X=KPZ, the dynamic scaling 
hypothesis leads to the KPZ equation \cite{KPZ86} 
for the RD+KPZ mix \cite{KNV06}:
\begin{equation}
\label{data}
h_t = \nu_0 f(p) h_{xx} + (\lambda_0 /2) f^{3/2}(p) h_x^2 + \eta (x,t) \, ,
\end{equation}
where $h \equiv h(x,t)$ is the height field, $x$ and $t$ are the spatial 
and time coordinates, respectively, subscripts denote partial 
derivatives, $\eta (x,t)$ is the white noise, and   
$\nu_0$ and $\lambda_0$ are constants. When $\lambda_0=0$, Eq. (\ref{data}) is 
the Edwards-Wilkinson (EW) equation \cite{EW82} when X=EW. 
When $\nu_0=\lambda_0=0$, Eq. (\ref{data}) defines universal 
RD dynamics.

Many simulation models of RD+EW and RD+KPZ growth 
processes \cite{KNV06,HA03,HA06,KN10} suggest $g(p)=p^\delta$ in 
Eq. (\ref{scaling}), which leads to the Family-Vicsek universal 
scaling \cite{FV85} of the average surface width $w(p,t)$ \cite{KNV06}: 
\begin{equation}
\label{FV}
w(p,t) = \frac{L^\alpha}{p^\delta} F \left( p^{2\delta} \frac{t}{L^z}\right) \, .
\end{equation}
For substrates of size $L$, $F(y)$ describes two limit regimes of evolution: 
$F(y) \sim y^{\alpha /z}$ if $y \ll 1$ (growth); and, $F(y) \sim \textrm{const}$ 
if $y \gg 1$ (saturation). In Eq. (\ref{FV}), $\alpha$ and $z$ are the universal 
roughness and dynamic exponents, respectively, of the universality class 
of the correlated-growth process X. The scale-dilatation 
exponent $\delta$ in the scaling  
prefactors in Eq. (\ref{FV}), however, is {\it nonuniversal}. It has been observed 
that, in some models, $\delta \approx 1$ across universality classes and, in  
some other models, $0<\delta \lessapprox 1$ within a single universality 
class \cite{KNV04,KNV06,Reis06}. Also, there are some 
models where the prefactors in Eq. (\ref{FV}) do not at all obey a power 
law in $p$ \cite{KN10,note02}. In the next section, 
we shall establish that this variation is 
not accidental in flux-conserving models 
but rather reflects the properties of the bulk of the 
deposited material.

\section{\textit{Ab initio} continuum equation 
by simplectic decomposition \label{conteq}}

Consider aggregations where identical particles fall onto a  
substrate of $L$ sites. On the substrate, the incoming particles  
may be accepted in accordance with a rule that generates correlations 
among the sites, i.e., in accordance to process X.  
It is understood here that the flux of the incoming particles 
is uniform and time independent, i.e., the average rate 
at which the particles arrive at the substrate does not 
vary with time and does not vary with position along the substrate. 
The correlated growth X occurs with probability $p$  
and competes with RD growth that occurs with probability $q=1-p$. 
It is understood here that the probability $p$ remains constant 
for the entire duration of the process RD+X, i.e., the average 
frequency of process X does not change with time and does not 
depend on the position of the site on the substrate.
When a particle is accepted at a site, 
the site increases its height by $\Delta h$. If, e.g., component {\it 1} is 
RD, and component {\it 2} is a correlated growth in the KPZ 
universality class, their corresponding growth equations are
\begin{eqnarray}
h_{1, t} &=& \eta_1 (x ,t) \, ,  \label{comp-1} \\
h_{2, t} &=& \nu_0 h_{2, xx} + (\lambda_0 /2) h_{2, x} ^2 + 
\eta_2 (x,t) \, ,  \label{comp-2} 
\end{eqnarray}
where $h_n(x,t)$, $n=1,2$, is the column height at $x$ after time $t$ 
when the component {\it n} acts alone. Assume for simplicity that the 
noise terms are of the same strength, i.e., $\eta \equiv \eta_1 = \eta_2$. 
In two-component growth, when both components act simultaneously together, 
the column height $h(x,t)$ is incremented due to either of the components 
with their corresponding probabilities $\tilde{p}$ 
and $\tilde{q}$, $\tilde{p}+\tilde{q}=1$:
\begin{equation}
\label{simplex}
\Delta h(x,t) = \tilde{p} \Delta h_2 (x,t) + \tilde{q} \Delta h_1 (x,t) \, .
\end{equation}
Here, probability $\tilde{p}$ (or $\tilde{q}$) is the fraction of contributions 
to $h$ from component {\it 2} (or {\it 1}). For some processes this fraction 
is identical to a fraction of times when $h(x)$ is incremented due to 
component {\it 2} (or {\it 1}) for the times from $0$ to $t$. However, 
as explained later, this is not so for all processes. 
In Eq. (\ref{simplex}), $\Delta h_n$ is understood as 
``being incremented due to the process {\it n},'' $n=1,2$. In this 
statistical sense, Eq. (\ref{simplex}) expresses a simplectic 
decomposition of $\Delta h(x,t)$ into its vertex components 
$\Delta h_n(x,t)$. Dividing Eq. (\ref{simplex}) by $\Delta t$  
and taking the limit $\Delta t \to 0$  gives the equation for 
time rates, $h_t = \tilde{p} h_{2,t} + \tilde{q} h_{1,t}$, 
into which we substitute Eqs. (\ref{comp-1}) and (\ref{comp-2}) 
to obtain: 
\begin{equation}
\label{combine}
h_t = \nu_0 \tilde{p} h_{2,xx} + (\lambda_0/2) \tilde{p} h_{2,x}^2
+ (\tilde{p}+\tilde{q}) \eta (x,t) \, .
\end{equation}
In Eq. (\ref{combine}), $h(x,t)$ is the column height that rises at $x$ as the result 
of two processes acting simultaneously from the beginning 
to time $t$. Here, $h_2(x,t)$ is the part of $h(x,t)$ that was 
created by component {\it 2} in this time. The other part was 
created by component {\it 1}. In other words, $h_2(x,t)$ is so 
far an unknown fraction of $h(x,t)$. To find a 
relation between $h$ and $h_2$, one must consider nonuniversal 
properties of aggregation processes.

We distinguish between the following two groups of surface growth processes. 
In one group we place all simple adsorption processes with conserved flux 
that do not create voids in the bulk of the deposited material. We call this group 
{\it adsorption-bulk-compact} (ABC) growths. For example, a simple  
random deposition or random deposition with surface relaxation fall into 
the ABC category. The other group, which we call 
{\it dense-or-lace-bulk} (DOLB) growths, contains 
processes that are not ABC-type. The DOLB group includes desorption 
processes that may lead to a dense bulk as well as adsorptions 
that lead to the formation of voids. 
The only type of desorption processes studied here are ones due to local 
spontaneous desorption at the surface, not desorption processes where an 
incoming particle strikes the surface and causes desorption.  
The latter type of desorption process would have shadowing effects, and hence 
would be extremely dependent on the direction of the incoming particles.  
Note that the DOLB category contains flux-conserving as well as 
flux-nonconserving processes. For example, ballistic deposition 
and deposition to local interface minima are both in the DOLB group.
Note that all RD universal processes are  
ABC-growth processes. As we show in the next paragraph, when component {\it 2} 
is of the ABC type, probabilities $\tilde{p}$ and $\tilde{q}$ in 
Eq. (\ref{simplex}) express fractional 
contributions to $h$ in terms of times, and then $h_2(x,t)=ph(x,t)$. 
This is not true when component {\it 2} is a DOLB growth.

Consider a discrete representation of events at coordinate $x$. 
Suppose there are $t$ deposition events in total, with $t_1$ 
events due to component {\it 1}, and $t_2$ events due to component 
{\it 2}, $t=t_1+t_2$. In ABC growth, after $t$ events, the total 
column height is $h=t \Delta h$, where contributions from components 
{\it 1} and {\it 2} are, respectively, $h_1=t_1 \Delta h$ and 
$h_2=t_2 \Delta h$. Thus, $h_1/h=t_1/t=q$ and $h_2/h=t_2/t=p$. 
Therefore, in ABC growth $h_2=ph$, and in 
Eqs. (\ref{simplex}) and (\ref{combine}) we can identify 
$\tilde{p}=p$ and $\tilde{q}=q$.

\begin{figure}[b]
\includegraphics[width=8.0cm]{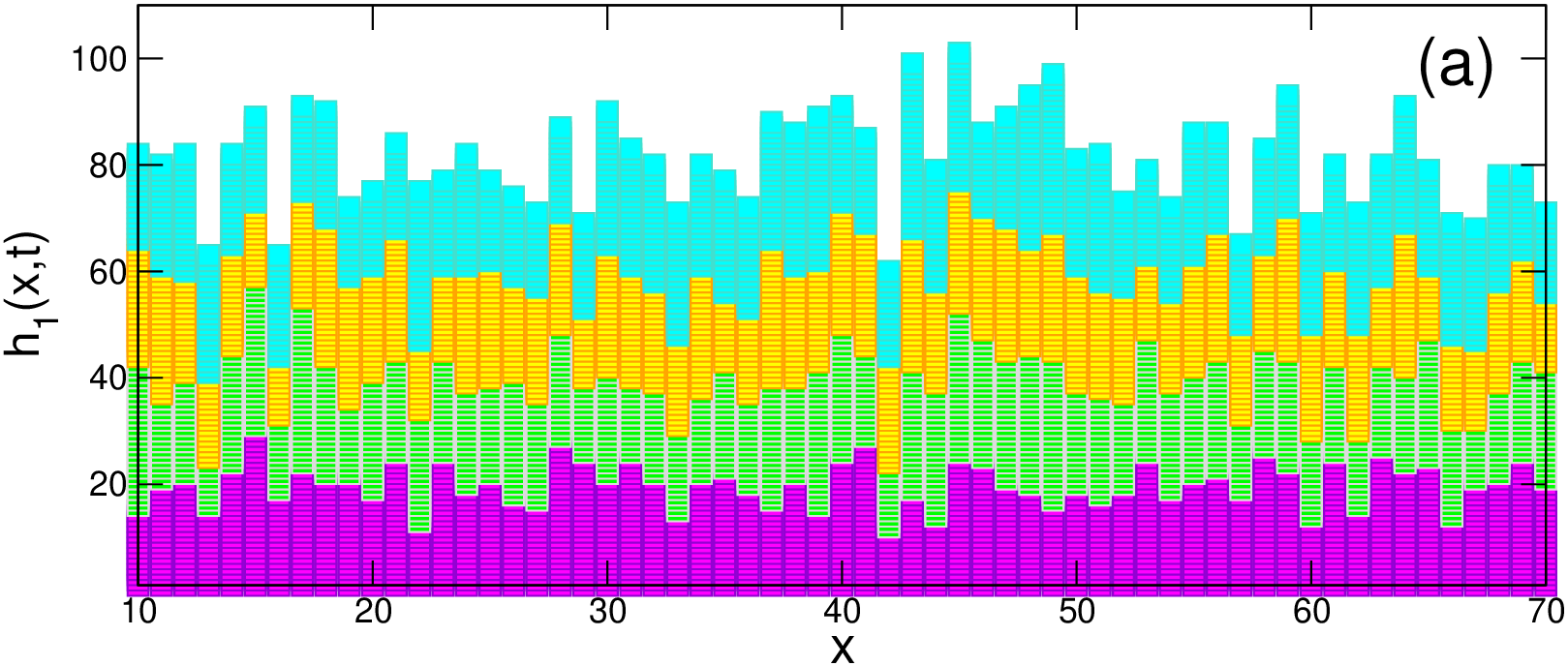}
\includegraphics[width=8.0cm]{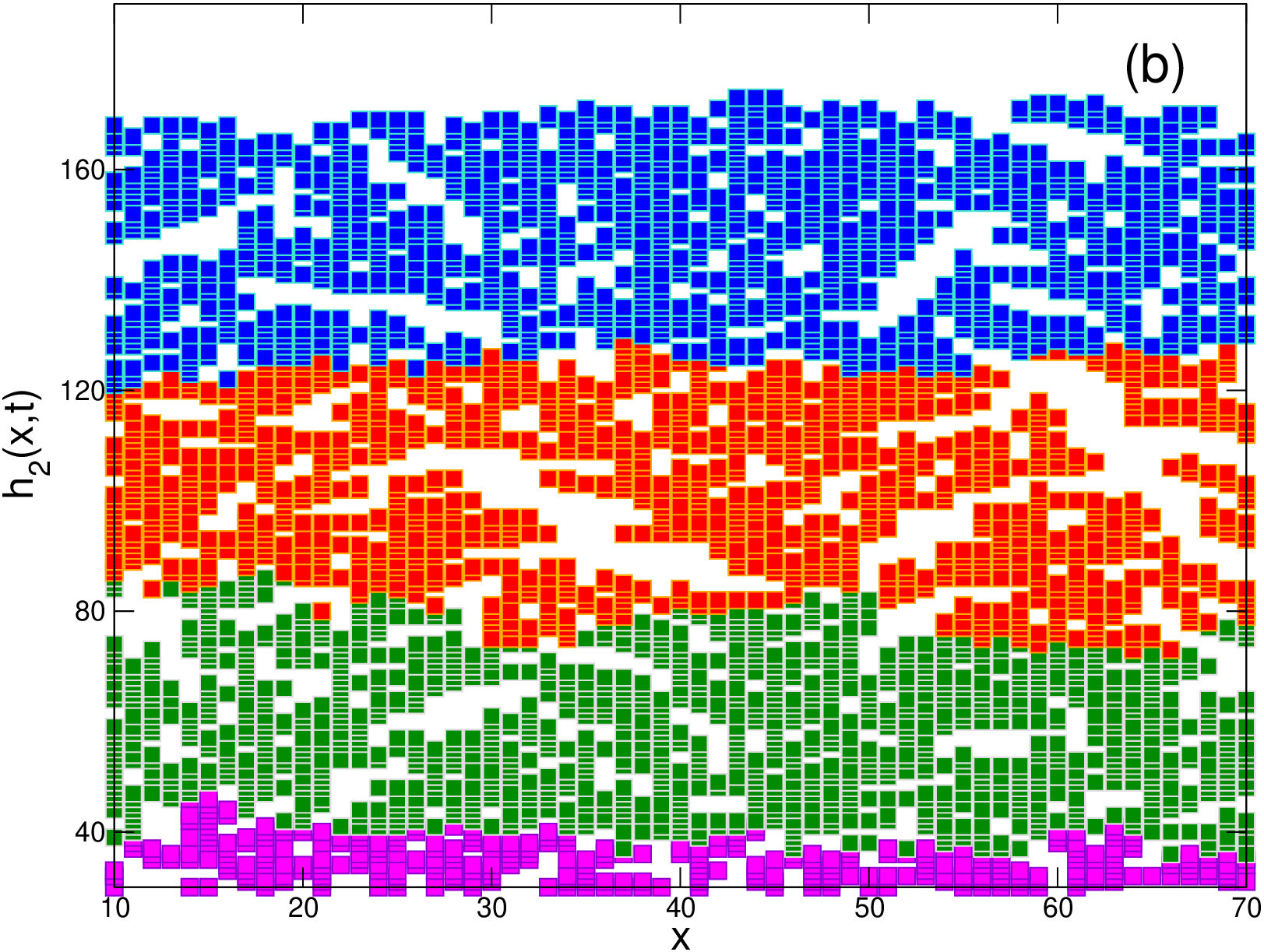}
\includegraphics[width=8.0cm]{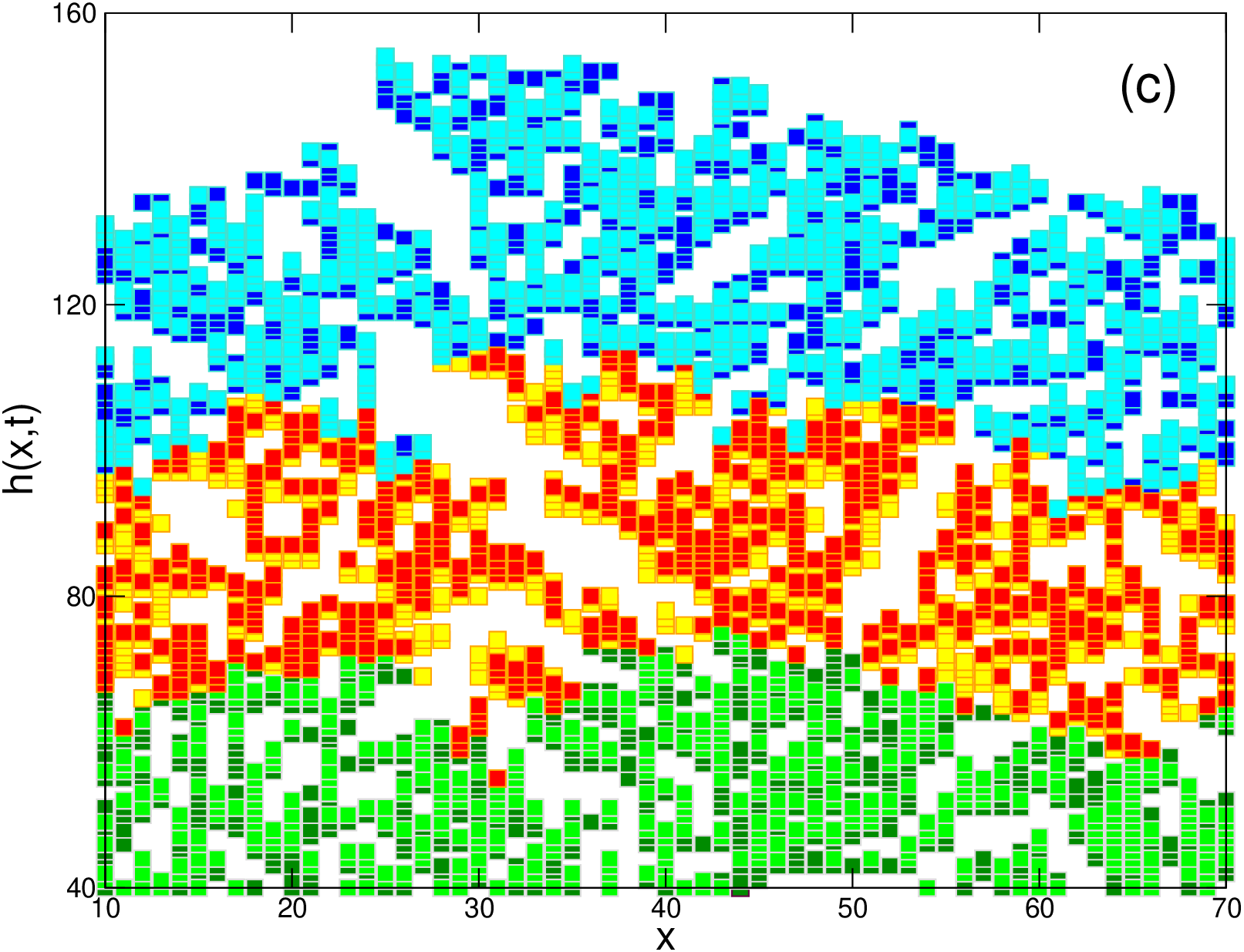}
\caption{\label{bdrd}
(Color on line) 
Bulk sections obtained in Monte Carlo simulations of surfaces, generated by: 
(a) random deposition (RD); 
(b) ballistic deposition (BD) that creates the bulk compactness $c=0.468$; and, 
(c) the competitive RD+BD process when $p=q=1/2$.
Coloring indicates time intervals. In panel (c), within a time interval, 
coloring is used to differentiate between deposits created by 
RD and those created by BD.
}
\end{figure}

Next consider that the component {\it 2} is a DOLB growth that creates voids. 
Now, an individual deposition event due to component {\it 2} not 
only increases $h$ by $\Delta h$, but may also result in the 
creation of voids 
[as illustrated in Fig. \ref{bdrd}{\it b}; an example of ballistic 
deposition]. The net effect is as though component {\it 2} 
deposited $\Delta h$ {\it and} the voids. Therefore, in $t_2$ events, 
its contribution to the column height is $h_2=(t_2+m) \Delta h$, 
where $m \Delta h$ reflects the increase in the field height due to the 
presence of voids. The component {\it 1} is RD, i.e., ABC-type, and 
$h_1=t_1 \Delta h$. After $t$ events, the net column height 
is $h=h_1+h_2=(t+m) \Delta h$. Thus, $h_1/h=t_1/(t+m) < t_1/t =q$ 
and $h_2/h=(t_2+m)/(t+m) > t_2/t = p$. The explicit form of 
these mutually complementary fractions, $h_n/h$ for $n=1,2$, allows them 
to be directly measured from the bulk. They are, in fact, 
the {\it effective probabilities} 
\begin{equation}
\label{effpq}
q_{\mathrm{eff}} \equiv h_1/h \quad \mathrm{and} \quad 
p_{\mathrm{eff}} \equiv h_2/h
\end{equation}
of deposition events due 
to components {\it 1} and {\it 2}, respectively, as  
would result from measuring the column height.

For some types of two-component growth with RD, the 
probability $p_{\mathrm{eff}}$ can be expressed approximately 
as the power law $p_{\mathrm{eff}}=p^\delta$ \cite{note02}, where the 
``best'' exponent $\delta$ can be estimated heuristically. 
For DOLB-type growth processes that produce voids, the exponent is 
$\delta <1$ because $p_{\mathrm{eff}} > p$. 
When the component {\it 2} is a DOLB growth with desorption, in the 
above reasoning one should change $m \to -m$. This will give 
$q_{\mathrm{eff}} > q$ and $p_{\mathrm{eff}}<p$, and 
$p_{\mathrm{eff}}=p^\delta$ with $\delta >1$. 
The value of $\delta$ depends on  nonuniversal particulars of the deposition rule of 
the component process {\it 2}. Therefore, in a DOLB growth $h_2=p_{\mathrm{eff}}h$, 
and in Eqs. (\ref{simplex}) and (\ref{combine}) 
$\tilde{p}=p_{\mathrm{eff}}$ and $\tilde{q}=q_{\mathrm{eff}}$.

In general, relations $h_2 (x,t)= p_{\mathrm{eff}}h(x,t)$ and 
$\tilde{p} \equiv p_{\mathrm{eff}}$ hold for all processes. 
When the correlation component is an ABC growth, its effective probability 
is identical with its frequency: $p_{\mathrm{eff}}=p$, 
provided that column-height increments are identical for the both processes 1 and 2. 
Thus, when process X is in the KPZ universality class, Eq. (\ref{combine}) 
gives the {\it exact} stochastic dynamics 
for the competitive RD+X processes:
\begin{equation}
\label{resulteq}
h_t = \nu_0 p_{\mathrm{eff}}^2 h_{xx} + (\lambda_0/2) p_{\mathrm{eff}}^3 h_{x}^2
+ \eta (x,t) \, .
\end{equation}

When the correlation component X is 
a DOLB growth, and when the effective probability is well approximated by 
a power law $p^\delta$, the above result can be summarized as 
$p_{\mathrm{eff}}=p^\delta$, 
where $\delta=1$ for ABC growths, and 
$\delta \ne 1$ for DOLB growths. This result is combined with 
Eq. (\ref{resulteq}) to give the {\it approximate} continuum equation 
for the RD+KPZ mix:
\begin{equation}
\label{RD_KPZ}
h_t=\nu_0 p^{2\delta} h_{xx} + (\lambda_0 /2) p^{3\delta} h_x^2 + \eta (x,t) \, .
\end{equation}

When in Eq. (\ref{comp-2}) $\lambda _0 \equiv 0$, the analogous 
reasoning gives the {\it exact} result for RD+EW dynamics:
\begin{equation}
\label{RD_EW}
h_t=\nu_0 p_{\mathrm{eff}}^2 h_{xx} + \eta (x,t) \, .
\end{equation}
In Eq. (\ref{RD_EW}), we can explicitly set $p_{\mathrm{eff}}=p$ because 
all processes in the Edwards-Wilkinson universality class are ABC-type 
processes. When the flux particles are identical, the exponent 
$\delta =1$ is exact.

\section{Discussion \label{discuss}}

Both results, Eqs. (\ref{resulteq}), (\ref{RD_KPZ}) and Eq. (\ref{RD_EW}), 
are in accord with 
our former derivation  that lead to Eq. (\ref{data}) \cite{KNV06}. 
Matching Eq. (\ref{RD_KPZ}) with Eq. (\ref{data})  
gives $f(p)=p^{2\delta}$, which form of $f(p)$ was used formerly  
to derive the approximate prefactors in Eq. (\ref{FV}). 
The inverse of the scaling (\ref{scaling}) when applied to 
Eqs. (\ref{RD_KPZ}) and (\ref{RD_EW}) transforms them to 
continuum equations for a ``pure'' correlated processes of $p=1$. 
Explicitly, it collapses {\it all} evolution curves $w(p,t)$ 
(for all $L$ and $p$) either onto $w(1,t)$ or onto a neighborhood 
of $w(1,t)$ \cite{KNV06}, following Eq. (\ref{FV}), 
provided the effective probabilities $p_{\mathrm{eff}}$ can be well 
approximated by the power-law $p^{\delta}$. When such a fit is not 
possible, Eq. (\ref{FV}) is still obeyed but then the scaling prefactors 
must be expressed directly in terms of effective probabilities. 
This is because the factor $p^\delta$ in the coefficients of 
Eq. (\ref{RD_KPZ}) is only a fit to the 
effective probability $p_{\mathrm{eff}}$.

\begin{figure}[t]
\includegraphics[width=8.0cm]{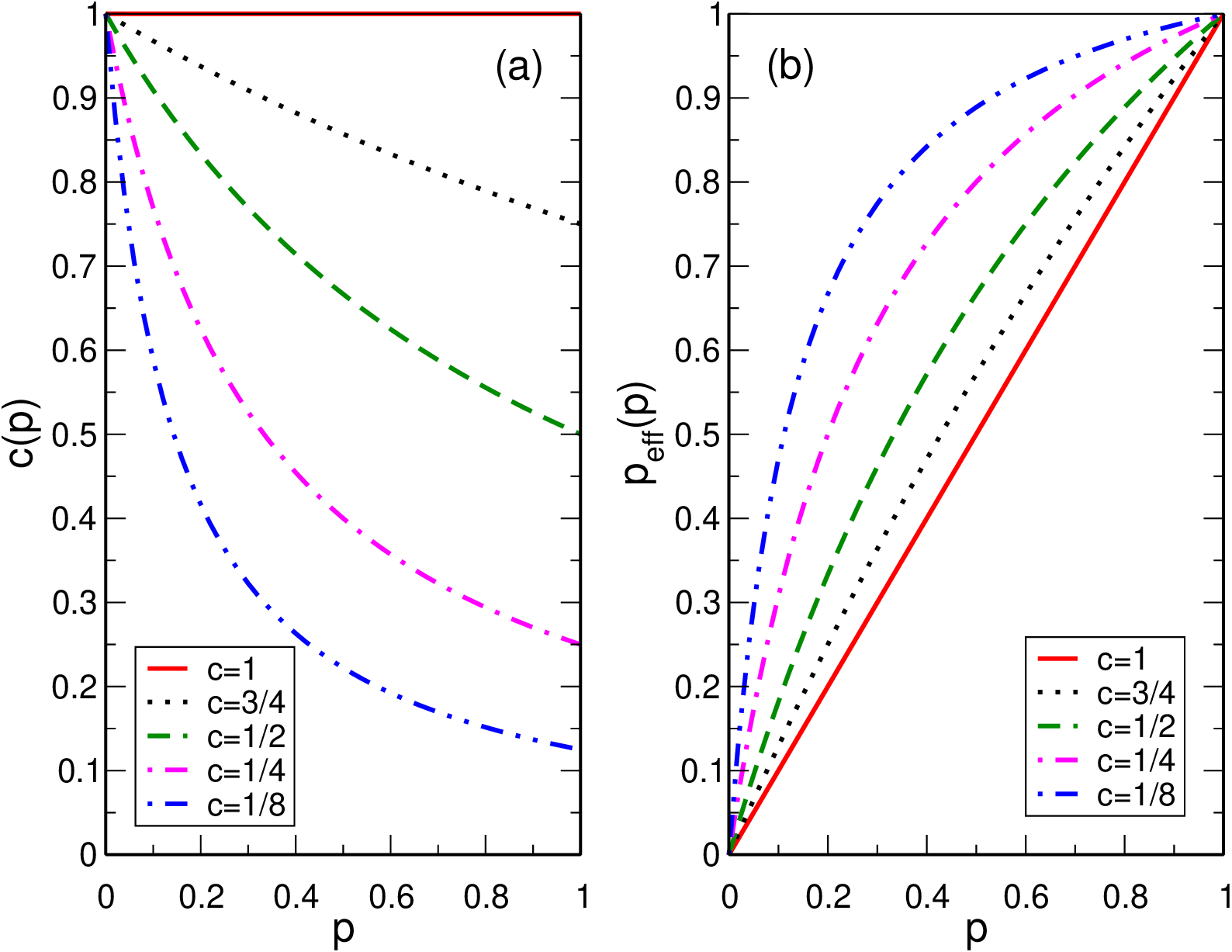}
\caption{\label{effective}
(Color on line)
Properties of the bulk formed in competitive process RD+X,
where X is an adsorption process that creates voids. When X acts alone,
i.e., in the absence of RD, it produces a bulk with compactness $c$.
(a) Compactness of the bulk $c(p)$ as a function of the activation
probability (i.e., frequency) $p$ of X, plotted for selected values of $c$.
(b) The effective probability
$p_{\mathrm{eff}}(p)$ of X for selected values of $c$.
}
\end{figure}

Effective probabilities, defined by Eq. (\ref{effpq}), 
are functions of the activation probability $p$ of process X, 
i.e., the average frequency of process X in the mix RD+X. 
Properties of the bulk morphology created by RD+X can be equivalently 
expressed either in terms of $p_{\mathrm{eff}}(p)$ or 
in terms of $q_{\mathrm{eff}}(q)$, because of the identities 
$p_{\mathrm{eff}}+q_{\mathrm{eff}}=1$ and $p+q=1$. 
Effective probability $p_{\mathrm{eff}}$ of X in the mix RD+X can be expressed 
in various equivalent functional forms that may involve either the average 
compactness $c(p)$ [or the number density of voids $v(p)$] of the bulk created 
by the mix RD+X or the average compactness $c$ [or the number density $v$ of voids]  
of the bulk created
by the process X acting alone (i.e., in the absence of RD when $p=1$):
\begin{equation}
\label{peff_1}
p_{\mathrm{eff}}(p)=\frac{h_2}{h}=\frac{m+t_2}{m+t}=
v(p)+p c(p) =1-q c(p),
\end{equation}
where $c(p)=t/(m+t)$, $v(p)=m/(m+t)$, and $c(p)+v(p)=1$;
\begin{equation}
\label{peff_2}
p_{\mathrm{eff}}(p)=\frac{h_2}{h}=\frac{m+t_2}{m+t}=
\frac{p}{c+pv} = \frac{p}{p+qc},
\end{equation}
where $c=t_2/h_2$, $v = m/h_2$, $p=t_2/t$, and $c+v =1$.
Combining Eq. (\ref{peff_1}) with Eq. (\ref{peff_2}) gives the compactness 
$c(p)$ of the RD+X bulk as a function of the activation probability $p$ 
of process X:
\begin{equation}
\label{comp}
c(p)=\frac{c}{c+pv}. 
\end{equation}
Equations (\ref{peff_1}), (\ref{peff_2}), and (\ref{comp}) show that 
$c(p)$ and $p_{\mathrm{eff}}(p)$ can be easily measured in experiment as well as 
in simulations: All that is needed is to measure the average density of 
voids in a sample cross section of the bulk when the correlated process X 
acts alone (i.e., in the absence of RD when $p=1$). 
Equations (\ref{peff_2}) and (\ref{comp}), plotted   
for several values of $c$ in Fig. \ref{effective}, 
show that neither the compactness $c(p)$ nor 
the effective probability $p_{\mathrm{eff}}(p)$ follow a power law 
in $p$ when $c<1$.

The dynamic scaling hypothesis for RD+X processes, Eq. (\ref{scaling}), 
can be reinstated by expressing $f(p)$ explicitly in terms of $p_{\mathrm{eff}}$: 
$f(p_{\mathrm{eff}})=g^2(p_{\mathrm{eff}})$, since Eq. (\ref{peff_2}) can be inverted 
to give $p(p_{\mathrm{eff}})$. 
Repeating the steps outlined in Ref. \cite{KNV06} gives the  
following generalization of Family-Vicsek scaling for the surface roughness:
\begin{equation}
\label{FVnew}
w^2(p,t)=\frac{L^{2 \alpha}}{f(p_{\mathrm{eff}})}
F_{\mathrm{RD+X}}\left(f(p_{\mathrm{eff}})\frac{t}{L^z}\right) \, ,
\end{equation}
where $F_{\mathrm{RD+X}}(\cdot)$ describes the three regimes of the evolution seen in 
Fig. \ref{geometric_scaling}b. The effect of the nonuniversal prefactors $f(p)$ in 
Eq. (\ref{FVnew}) is a dilatation of length and time scales, as discussed in 
Ref. \cite{KNV06}. The physical meaning of $p$ is that of a noise-tuning parameter.

\begin{figure}[b]
\includegraphics[width=8.0cm]{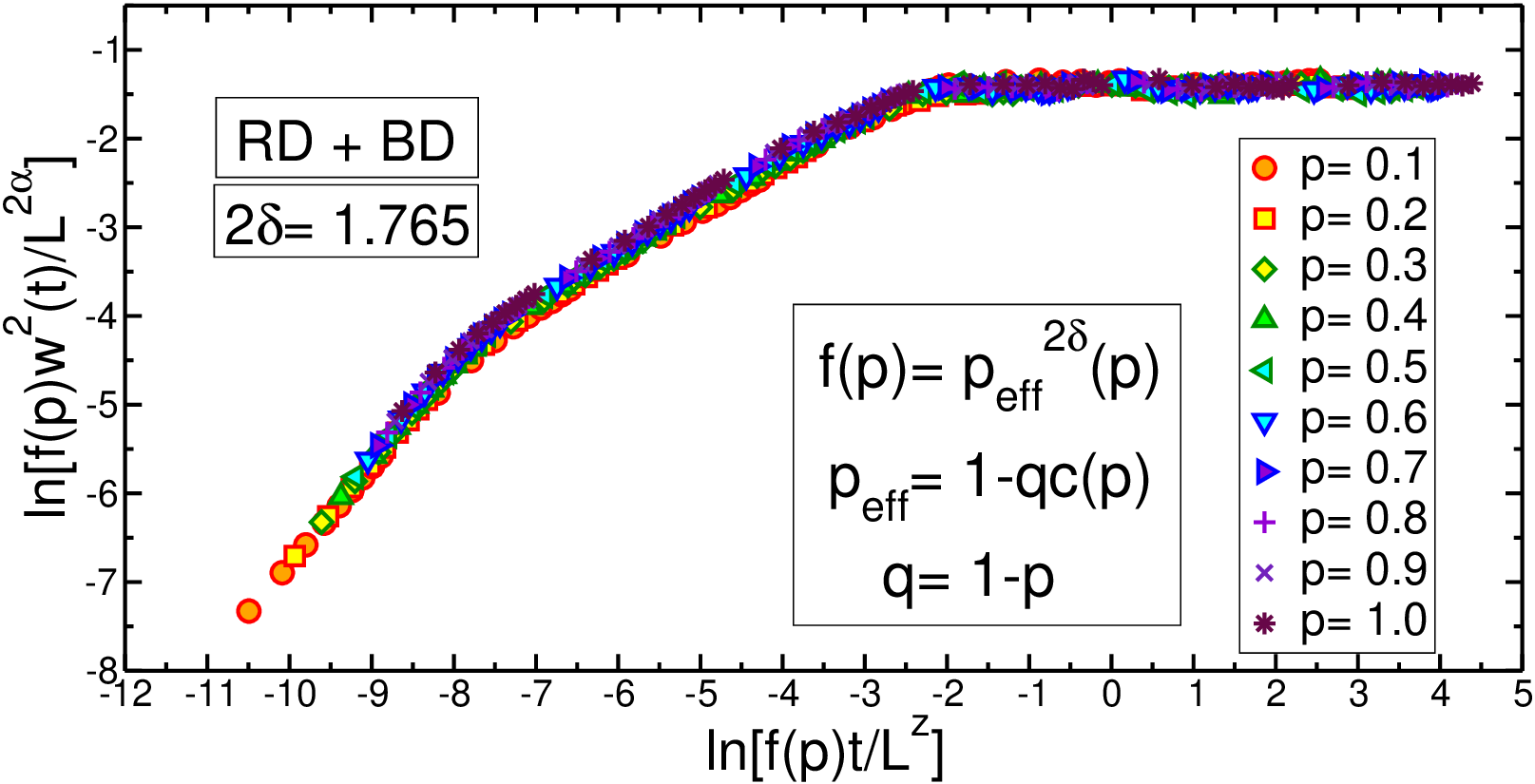}
\caption{\label{effective_scaling}
(Color on line)
Scaled time-evolution $w^2(p,t)$ in the RD+BD model. In this example,
the scaling function $f(p)=g^2(p)$ explicitly incorporates the compactness
$c(p)$ of the bulk formed in the RD+BD process. Here, $L=500$,
$2\alpha=1$, and averaging was performed over 400 surface configurations, 
i.e., independent simulations.
}
\end{figure}

In Fig. \ref{effective_scaling} 
we give an example of the exact scaling where nonuniversal 
prefactors in Eq. (\ref{FV}) are directly expressed by $p_{\mathrm{eff}}$ 
via the substitution $p^{\delta} \to p_{\mathrm{eff}}^{\delta}(p) = \sqrt{f(p)}$ 
for the RD+BD model when ballistic deposition (BD) is the nearest-neighbor (NN) sticking 
rule \cite{BS95}. Here, the effective probability depends on both $p$ and 
the mean compactness $c(p)$ of the bulk formed in the RD+BD process, given 
by Eq. (\ref{peff_1}) \cite{note02}. The excellent data collapse in the 
\textit{full range} of $p\in (0;1]$, seen in Fig. \ref{effective_scaling}, 
can be contrasted with Fig. 5 of Ref. \cite{KNV06} that shows only an 
approximate data collapse for the same system with the best fit exponent 
$\delta \approx 0.41$ in Eq. (\ref{FV}). It needs to be said explicitly 
that the scaling where $\delta=1/2$ in Eq. (\ref{FV}), proposed in 
Refs. \cite{HA06,BL05} for RD+BD models, does not produce data 
collapse at all. 
The RD+BD model when BD is the next-nearest-neighbor (NNN) sticking rule \cite{BS95} 
provides an example where $p_{\mathrm{eff}}(p)$, 
and thus the nonuniversal prefactors $f(p)$ and $g(p)$  
in Family-Vicsek universal scaling, cannot be expressed by 
a power law $p^\delta$ \cite{KN10}. In this system the surface roughening 
obeys power laws in effective probability that incorporates either 
the compactness or the void density of the bulk, resulting in excellent 
data collapse of $w(p,t)$, similar to that 
seen in Fig. \ref{effective_scaling} \cite{note02}.

The approach introduced here by the example of a KPZ processes, 
can be applied to a broad range of stochastic growth models RD+X, where 
component {\it 2} can be any isotropic growth in $(1+n)$ dimensions:
\begin{equation}
\label{process-2}
h_{2,t}(\vec{x},t)= \hat{D}(h_2)  + \eta_2 (\vec{x},t) \, ,
\end{equation}
where $\vec{x}$ is $n$ dimensional, and the operator $\hat{D}$ represents 
only local interactions \cite{BS95}. 
In the general case, Eq. (\ref{combine}) is written as  
$h_t=p_{\mathrm{eff}} h_{2,t} + q_{\mathrm{eff}} h_{1,t}$, and  
combined with Eqs. (\ref{comp-1}) and 
(\ref{process-2}) to find for the competitive growth
\begin{equation}
\label{competitive}
h_t (\vec{x},t) = 
p_{\mathrm{eff}} \hat{D}(p_{\mathrm{eff}} h)  + \eta (\vec{x},t) \, ,
\end{equation}
where $\eta = (1-p_{\mathrm{eff}}) \eta_1 + p_{\mathrm{eff}} \eta_2$, 
and the noise strengths may be different.   
Equations (\ref{process-2}) and (\ref{competitive}) represent the same  
universality class since the multiplication by $p_{\mathrm{eff}}$ 
does not modify local interactions:  $p_{\mathrm{eff}}$ affects the noise 
strength and the gradient of the height field, but does not generate any new terms 
other than those already given by operator $\hat{D}$. Hence, if a correlated growth 
belongs to a given universality class, 
its mix with RD will remain in the same universality class. 
Elementary calculations show that Eq. (\ref{competitive}) is invariant 
under the scaling $g(p) h(\vec{x},t)=h'(\vec{x},t'=f(p)t)$. 
If $g(p)=p_{\mathrm{eff}}(p)$ and $f(p)=p_{\mathrm{eff}}^2 (p)$, and if 
the noise strengths are the same, this scaling maps the universal dynamics 
(\ref{competitive}) of RD+X onto the universal dynamics of X. 
In this case the invariance implies $g(p)w(p,t)=w'(f(p)t)$, 
where $w'(\cdot)$ has universal scaling properties of the process X. 
When X is either in the KPZ or in the EW universality 
class, and if additionally $p_{\mathrm{eff}}\approx p^{\delta}$, we 
recover Eq. (\ref{FV}).

When both the RD and the correlation component {\it 2} 
have deposits of unit height, when $p_{\mathrm{eff}}\approx p^{\delta}$, 
we have $\delta =1$ if component {\it 2} is of the ABC-type; 
and, $\delta \ne 1$ if it 
is of the DOLB type. In the latter case, the value of the exponent 
$\delta$ is specific to component {\it 2}. 
When $p_{\mathrm{eff}}$ incorporates explicitly bulk properties, 
the scaling is $g(p)=p^{\delta}_{\mathrm{eff}}(p)$, 
where the new scale-dilatation exponent $\delta$ is obtained from 
the slope of $\ln{w^2(p)}$ plotted vs $\ln{p_{\mathrm{eff}}(p)}$ at saturation. 
In DOLB growth with voids, $p_{\mathrm{eff}}$ can be determined 
by measuring the mean density of voids in the bulk 
(Figs. \ref{bdrd} and \ref{effective_scaling}). 
Similarly, in DOLB growth with desorption, $p_{\mathrm{eff}}$ is 
connected to the mean fraction of the removed material 
(or flux) \cite{note02}.

\begin{figure}[tp]
\includegraphics[width=8.0cm]{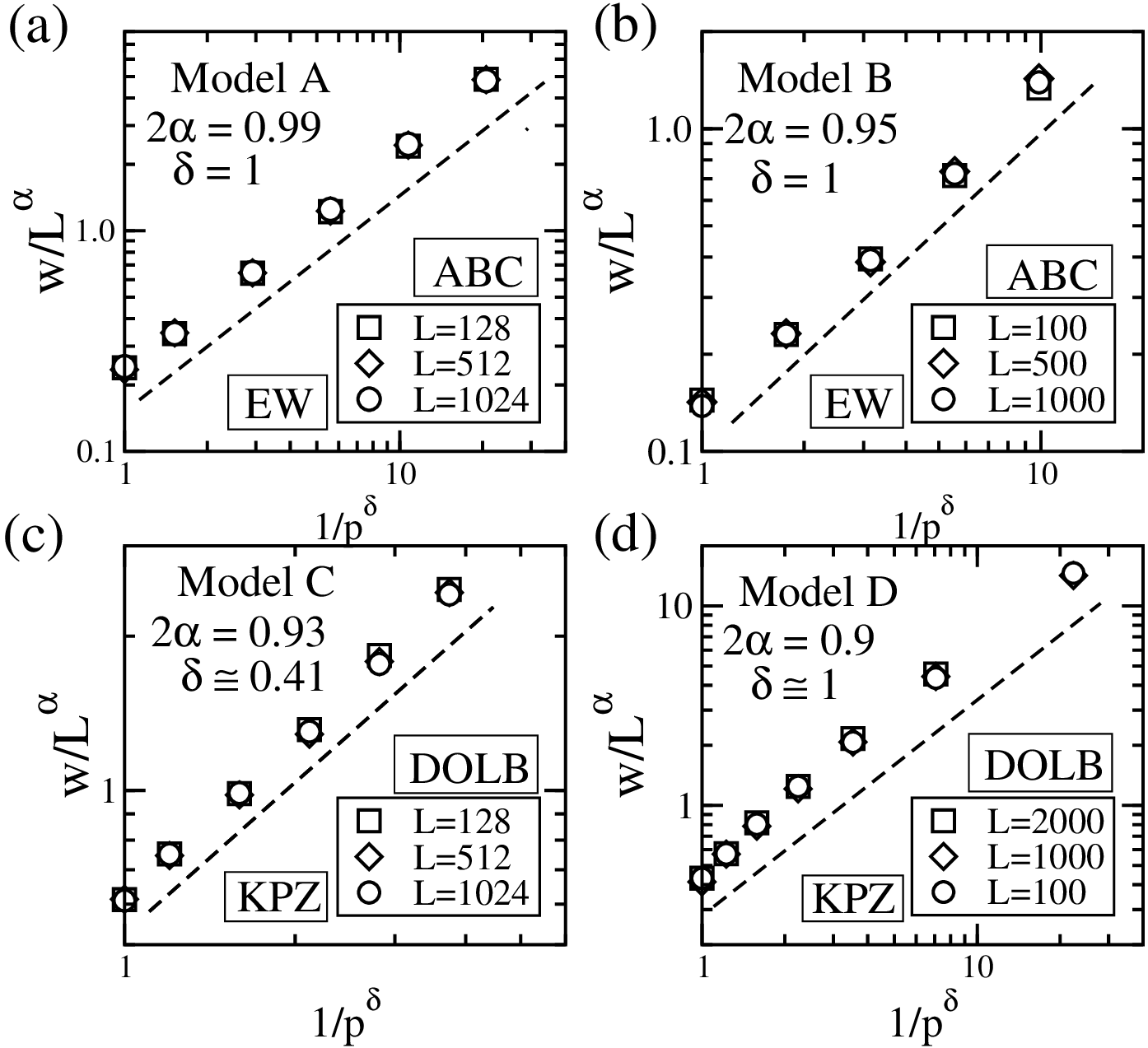}
\caption{\label{fig-1} 
Scaled widths at saturation $w$ vs the parameter
$1/p^{\delta}$. Panels (a) and (b) are for {\it models A} and {\it B}, respectively. 
Panels (c) and (d) are for {\it models C} and {\it D}, respectively.
Reference lines have slope $1$.
Data are scaled with the $\alpha$ values shown.
}
\end{figure}

The analysis presented here explains scaling results of the following  
mixed-growth models in $(1+1)$ dimensions.   
{\it Model A} \cite{HA03,HMA01,KNV06}:  
component {\it 2} is RD with surface relaxation. 
{\it Model B} \cite{KNV06}:  component {\it 2} simulates a deposition 
of a sticky nongranular material of variable droplet size.  
{\it Model C} \cite{HA03,HA01,KNV06}: 
component {\it 2} is the NN sticking rule of BD. 
{\it Model D} \cite{KNV06,KNV04}:  component {\it 2} is a 
deposition of Poisson-random numbers to the local surface minima. 
{\it Models A} and {\it B} are ABC growths in the EW 
universality class, where $p_{\mathrm{eff}}=p$ and $\delta=1$ 
(Fig. \ref{fig-1}). {\it Models C} and {\it D} belong to the KPZ universality 
class. {\it Model C} is an example of DOLB growth with voids, 
with a 53.2\% void density in the bulk when $p=1$, and in this case 
$\delta \approx 0.41 <1$. {\it Model D} is a DOLB-type growth 
that produces a compact bulk but component {\it 2} is flux nonconserving, 
and here $\delta \approx 1$. 
Extensions of {\it Models A} and {\it C} to ($1+n$) 
dimensions \cite{HA03}, $n=2,3$, yield results that conform to 
our theoretical predictions of 
$p_{\mathrm{eff}} \approx p^\delta$ with $\delta \ne 1$ for 
mixing RD with DOLB processes, and $\delta =1$ for 
mixing RD with ABC processes. Additional examples
include cases \cite{Reis06} when component {\it 2} is a restricted Kim-Kosterlitz 
solid-on-solid model \cite{KK89} (where RD+X is in KPZ universality class), and when 
it simulates a conserved restricted SOS growth of 
Kim \textit{et al.} \cite{Kim94}. In the latter case the process RD+X is in 
the Villain-Lai-Das Sarma universality class \cite{Reis04,Vil91,LD91}.

An interesting lattice simulation model has been recently considered 
by Banerjee {\it et al.} \cite{BSR14} in an attempt to describe a realistic 
sedimentation. The Banerjee {\it et al.} model is a competitive growth 
process that has three component processes: one RD process and two 
DOLB processes, where one DOLB process is BD with the NN sticking 
rule and the other DOLB process is BD with the NNN sticking rule. 
In the language of our study, the overall process is the RD+X process, 
where X=X1+X2 and, as the convex linear combination of two KPZ processes, 
X is in KPZ universality class. Accordingly, this system should obey the 
scaling law of Eq. (\ref{FVnew}) in the effective probability 
of the combined process X. 
Time-evolution plots of the surface roughness 
in Ref. \cite{BSR14} suggest such scaling.

The extension of the approach presented here to other competitive 
growth processes may provide a tool to understand the 
observed dynamics of surface growth. Realistic systems may involve 
many component processes, some of which may be dominant. Within our 
formalism a departure point may be a generalization of Eq. (\ref{simplex}):
\begin{equation}
\label{g-simplex}
\Delta h(\vec{x},t) = \sum_k p_{\mathrm{eff}}^{(k)} \Delta h^{(k)} (\vec{x},t) \, ,
\end{equation}
where the summation is over all contributing processes, and 
$ \Delta h^{(k)}$ is the column-height increment due to the $k{\rm th}$ process. 
In a first approximation, component processes are not explicitly correlated. 
Each process is encountered with the activation probability 
or frequency $p_k$, $\sum_k p_k=1$, and 
contributes to the growth with an effective probability $p_{\mathrm{eff}}^{(k)}$,   
$\sum_k p_{\mathrm{eff}}^{(k)} =1$. In the trivial case when all components are  
ABC-type models with the 
unit mean deposit height we have $p_{\mathrm{eff}}^{(k)}=p_k$. 
For a DOLB-type growth the $p_{\mathrm{eff}}^{(k)}$ will have to be determined. 
Depending on the model, $p_{\mathrm{eff}}^{(k)}$ 
can be estimated by analyzing the growth 
when process $k$ acts alone, and measuring either the mean bulk 
density or the mean fraction of the detached material or 
both \cite{note02}. Simplectic decompositions like the one proposed in 
Eq. (\ref{g-simplex}) have a long history of applications in many diverse fields 
and are the precursors of probability measures.

Stochastic theory of multicomponent competitive far-from-equilibrium 
surface-growth processes is a newly emerging topic in statistical physics. 
During the recent two decades, in addition to model-specific simulation 
studies of two-component (either RD+EW or RD+KPZ lattice) growth models, 
a special case of RD+RD has been considered both in a theoretical 
mean-field approach and in simulations \cite{NK09}. The absence 
of a consistent continuum theory for the RD+X mix hindered 
scale-invariance studies of more complex systems such as EW+EW 
or KPZ+KPZ or EW+KPZ, or more general three-component systems  
(such as, e.g., those in Ref. \cite{BSR14}). Understanding the dynamics 
of real physical systems calls for more realistic models that 
would go beyond a one-process theory of kinetic roughening. 
For example, in realistic modeling of ion-bombardment experiment, 
where shadowing effects matter, the mechanism of ion desorption 
must be (at least) accompanied by ion diffusion along the substrate 
as well as by random deposition. The construction of a continuum 
stochastic growth equation, outlined in this article, will be helpful for 
future studies of two- and three-component competitive nonequilibrium 
growth systems.

\section{Conclusion \label{conc}}

In summary, we derived continuum stochastic-growth equations 
and the resulting scaling for competitive RD+X growth processes.   
The RD+X growth processes 
show that model-dependent prefactors in universal scaling laws can be 
linked with the bulk morphology and 
determined from bulk structures. This necessitates the distinction 
between the adsorption-bulk-compact (ABC) and the dense-or-lace-bulk (DOLB) growth 
processes in dynamic-scaling analysis of competitive mixed-growth models.
For competitive systems, the activation probability of process X, i.e., 
its frequency in the RD+X mix, alone does not provide sufficient 
information to correctly describe their dynamics. The essential 
physical nonuniversal parameter here is the effective probability, 
$p_{\mathrm{eff}}$, of X.   The bulk 
morphology allows one to obtain $p_{\mathrm{eff}}$ for either 
experimental or computational studies.  
Furthermore, $p_{\mathrm{eff}}$ is parametrized by the activation probability. 

\begin{acknowledgments}
We thank F.D.\ A.\ Aar\~{a}o Reis and H.E.\ Stanley for stimulating correspondence. 
This work was partially supported by NSF (USA) Grant DMR-1206233, 
and by the HPC$^2$ CCS at Mississippi State University (USA). 
Resources of the National Energy Research Scientific Center, 
supported by the Office of Science of the DoE (USA) 
under Contract No. DE-AC03-76SF00098, are acknowledged.  
\end{acknowledgments}

\end{document}